\documentclass[preprint,review,12pt]{elsarticle}

\usepackage{amsmath}
\usepackage{amsfonts}
\usepackage{amssymb}
\usepackage{bm}
\usepackage{latexsym}
\usepackage{float}
\usepackage{graphicx}
\usepackage{color}
\usepackage{lineno}

\newcommand{\commentMP}[1]%
{\textsf{\textcolor{red}{#1$^{\mathrm{MP}}$}}}
\journal{Macromolecular Theory \& Simulations}

\begin{document}

\begin{frontmatter}

\title{Predictors of cavitation in glassy polymers under tensile
strain: a coarse grained molecular dynamics investigation}

\author[1,2]{Ali Makke}
\ead{ali.makke@insa-lyon.fr}
\author[2]{Michel Perez}
\author[3]{J\"org Rottler}
\author[2]{Olivier Lame}
\author[1]{Jean-Louis Barrat}

\address[1]{Universit\'e de Lyon- Univ. Lyon I - LPMCN - UMR CNRS 5586- F69622 Villeurbanne, France}
\address[2]{Universit\'e de Lyon - INSA Lyon - MATEIS - UMR CNRS 5510 - F69621 Villeurbanne, France}
\address[3]{Department of Physics and Astronomy, University of British Columbia, 6224 Agricultural Road, Vancouver, BC, V6T 1Z1, Canada}

\begin{abstract}
The nucleation of cavities in a homogenous polymer under tensile strain is investigated in a coarse-grained molecular dynamics simulation. In order to establish a causal relation between local microstructure and the onset of cavitation, a detailed analysis of some local properties is presented.  In contrast to common assumptions, the nucleation of a cavity is neither correlated to a local loss of density nor, to the stress at the atomic scale and nor to the chain ends density
in the undeformed state. Instead, a cavity in glassy polymers nucleates in regions that display a low bulk elastic modulus. This criterion allows one to predict the cavity position before the cavitation occurs. Even if the localization of a cavity is not directly predictable from the initial configuration, the elastically weak zones identified in the initial state emerge as favorite spots for cavity formation. 
\end{abstract}




\end{frontmatter}
\begin{linenumbers}

\section{Introduction}
\label{sec:intro}

Under hydrostatic stress conditions, failure of amorphous polymers occurs
through cavitation, often followed by crazing, \emph{i.e.} the
formation of interpenetrating micro-voids \cite{Perez98}.
Similarly, the plastic deformation of semi-crystalline polymers is
strongly correlated to the nucleation of cavities in the amorphous
region~\cite{Humbert10}.  Although essential to control deformation
and failure of many organic materials, cavitation in glassy polymers
under load is poorly understood. To our knowledge, the
microstructural causes, or the precursors of cavitation at a
microscopic scale, are not clearly identified.  Although it is known
that impurities or surface defects aid the nucleation of cavities ~\cite{Hermann02,Argon77}, it
is presently not possible to predict where cavitation will take place
in the polymer.

Classical nucleation theory, where elastic energy is balanced by the creation of free surface, was used by Argon to model the cavitation nucleation~\cite{Argon77}. 
Estevez \textit{et al.} investigated the fracture toughness in glassy polymers using mechanical approaches with empirical constitutive equations to describe the competition between shear yielding and crazing~\cite{Estevez00}. They noted that the development of crazes is favored by a fast local deformation. 

According to Gent \cite{Gent70}, crazing in glassy plastics can be attributed to a local stress-activated devitrification.  It is generally agreed that large triaxial tensile
stresses are needed to induce cavitation, which forms the basis of several macroscopic craze initiation and cavitation criteria \cite{Sternstein69, Bowden73}.  
Molecular dynamics simulations of polymer glasses also found a transition from shear yielding, which obeys a pressure-modified von Mises yield criterion \cite{Rottler01},
to cavitation as the hydrostatic pressure becomes negative, but have
not yet investigated the connection between cavitation and local
microstructural configuration. More recent simulations explored
correlations between the location of failure, higher mobility regions and a higher chain ends density~\cite{Sixou06} acting then as local defects, or a local, stress-induced disentanglement of chains \cite{Mahajan10}. In the latter work a primitive path algorithm was used to monitor
the entanglement network in a sample undergoing triaxial deformation, and it was found that regions undergoing crazing were also depleted in terms of entanglements.

 The local mechanical properties are a determining  factor to understand the response of systems under strain. Yoshimoto et al. \cite{Yoshimoto05,Yoshimoto04} have calculated the local elastic modulus in a coarse grained polymer glass using a thermodynamic approach based on stress fluctuation . They found that polymers are mechanically heterogenous at local scale.
Papakostantopoulos et al. \cite{Papakonstantopoulos08}  have studied the earliest local plastic events observed in the elastic regime of polymer glass. They found that these irreversible events take place in domains that exhibit a low positive elastic modulus.
Analogous results were obtained by Tsamados et al. on Lennard-Jones glasses submitted to a quasistatic shear strain\cite{Tsamados09}. They found a correlation between high nonaffine displacements and local low elastic modulus. 

It is unclear, however, if these criteria can be used in a predictive manner, in the sense that the cavitation event could be predicted from the configuration of an unstrained system.

In this paper, we will therefore investigate the correlation between
the microstructure of a homopolymer at the segmental level and the
nucleation of cavities, in an attempt to find a microstructural
\textit{ predictor} of such events. Section~\ref{MethodModels.sec}
will present the methodology and will demonstrate that the non-affine
particle displacement (NAD) is a particularly suitable tool for
characterizing and locating
cavities. Section~\ref{CausesOfCavitation.sec} will be devoted to the
investigation of possible causes of cavitation, namely (i) local
excess of free volume, (ii) local excess of atomic stress, (iii) local
density of beads and chain ends, and (iv) local bulk modulus. We will
show that while (i)-(iii) bear little correlation to the NAD, the
local bulk modulus (iv) has a much better potential to predict the
cavitation event.

\begin{figure*}
\begin{center}
\includegraphics[width=\textwidth]{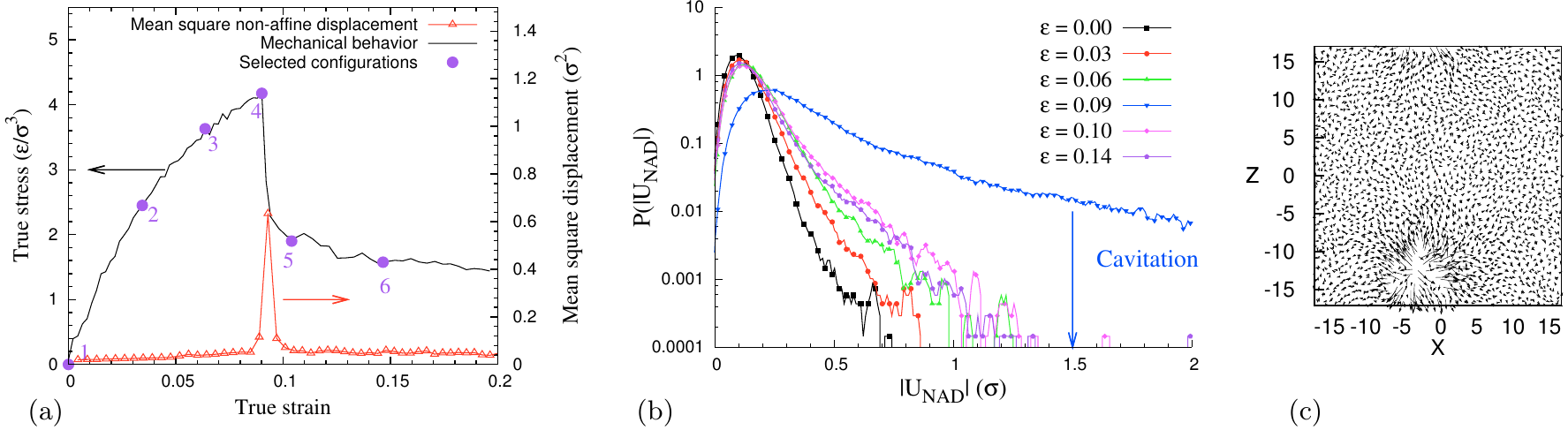}
\end{center}
\caption{(a) Stress-strain curve of a glassy polymer at $T=0.2$ during
  a triaxial tensile test. A peak in the mean square non-affine
  displacement is observed simultaneously with the drop of the stress
  due to cavitation. Markers 1, 2, 3 and 4 indicate the selected
  configurations for which distributions of NAD are plotted in
  (b). These distributions show an exponential tail for configuration
  4, due to the large amplitude motions caused by cavitation. (c)
  Representation of the NAD in a cross-section of the sample
  containing the cavity, for configuration 4.}
\label{naf_displacement.fig}
\end{figure*}

\section{Methods}
\label{MethodModels.sec}

\subsection{Molecular dynamics simulations}
Molecular dynamics (MD) simulations were carried out for a well
established coarse-grained model, in which the polymer is treated as a
linear chain of $N$ beads of mass $m$, which we refer to as monomers,
connected by stiff anharmonic springs that prevent chain crossing and
breaking \cite{Kremer90}. The beads interact through a conventional
6-12 Lennard-Jones potential that is truncated at 2.5 times the
particle diameter (The tiny discontinuity of the force at the cutoff distance, less than 1\% of the maximal attractive part, has no consequence). All
quantities will be reported in units of the Lennard-Jones length scale
$\sigma$ and energy scale $\epsilon$, and the characteristic time is
$\tau_{LJ}=\sqrt{\mathrm{m}\sigma^2/\epsilon}$. Newton's equations of
motion are integrated with the velocity Verlet method and a time step
$\Delta t=0.006$. Periodic simulation cells of initial size
$L_x(0)=L_y(0)=L_z(0)=34.2$ containing $M=215$ chains of size $N=200$
beads were used with a Nos\'{e}-Hoover thermostat, \emph{i.e.}  in the
NVT ensemble.
All samples were generated using the ``radical-like'' polymerization method \cite{Perez08}. The polymerization starts from a Lennard-Jones liquid, where 215 beads are chosen randomly to behave as ``radical'' sites. Each radical bead is allowed to connect to a free and nearest neighbor with a strong covalent bond. The radical sites are then transferred to the new connected beads, allowing thus the growth of all chains. If no monomers are near the radical, no FENE bond is created. Another attempt will be performed at the next growth stage. Between two growth stages, the entire system is relaxed during 100 MD steps. The polymerization propagates until all chains reach their target length of 200 beads. When the generation is terminated, residual single beads are removed and the system is relaxed for $10^7$~MD steps in NPT ensemble at $T=1$ and $P=0$ to reach an equilibrium state. The equilibration leads to a ``mean square internal distance'' very close to the function given by Auhl et al.~\cite{Auhl03}. The polymer is then rapidly quenched into the glassy state at a temperature $T=0.2$ in NPT ensemble (cooling rate: 1$\epsilon$ per $10^6$ MD steps). The glass transition temperature is $T_g=0.43$ \footnote{$T_g$ has been determined by the slope change observed when the sample volume is plotted with respect to the temperature during cooling from $T = 1$ to $T = 0.0001$ under the NPT ensemble}. The pressure remains zero and the sample density reaches 1.04 before applying the deformation.

Triaxial tensile test conditions were employed \cite{Makke09}. The
samples were subjected to a sequence of deformation-relaxation steps,
composed of (i) a rescaling of the simulation box in the tensile
direction ($y$ in our case so that the true strain
$\epsilon_{yy}(t)=\ln(L_y(t)/{L_y(0)}$), whereas the two other
dimensions remain unchanged, followed by (ii) an MD step in the NVT
ensemble.  The deformation rate was chosen to be $\dot{L}_y=0.0025$,
so that the initial strain rate is
$\dot{\epsilon}_{yy}(0)=7.3\times10^{-5}$.  Over the range of strain investigated, the 
true strain rate remains essentially constant. Note that the applied
deformation trajectory leads to a high level of triaxiality, which is
the basic ingredient for cavitation. As the deformation proceeds,
configurations were recorded along the trajectory in order to analyze
their microstructure.

\subsection{Non-affine displacement: a tool for characterizing cavitation}
\label{NonAffineDisplacement.sec}
The mechanical behavior of our glassy polymers under triaxial tensile
conditions is illustrated in Figure
\ref{naf_displacement.fig}(a). Three main regimes can be
distinguished: (i) elastic, (ii) viscoelastic, and (ii) drawing
regime, which occurs at constant stress. In the elastic regime, the
increase of deformation will slightly shift the bead positions from
their local energy minima, resulting in reversible behavior. This
regime is limited to a very low strain 0.001 as demonstrated by
Schnell~\cite{Schnell06}. In the viscoelastic regime, stress is
relaxed by inter-chain sliding. This stage is limited by a strong drop
of stress. When a critical deformation is reached, cavities will
nucleate and then part of the stored elastic energy is released as
free surfaces open up. Note that the strain hardening regime is not
shown in Figure \ref{naf_displacement.fig}(a), since it occurs at
larger strains when the entanglement network of chains and fibrils
becomes stretched \cite{Rottler01}.

The detection of cavity nucleation could be performed visually on
snapshots that are regularly stored during the course of the tensile
test. However, small cavities in a three dimensional sample can be
delicate to observe. Therefore, a more versatile indicator is
needed. The non-affine displacement (NAD) is the perfect candidate for
such observation and has been successfully used to monitor local
plastic activity in 2D amorphous Lennard-Jones packings under athermal
quasistatic deformation~\cite{Tsmamados10}. Note that NAD fluctuations
can not find their origin in the thermal motion of atoms since, in the
framework of this paper, specimens are maintained well below their
glass transition temperature ($T=0.2 < T_g=0.43$).  Moreover, the NAD
can be used as a routine tool and it starts to increase locally, in
the early stages of cavity nucleation, even before the cavity could be
observed visually on a snapshot of the sample.

The non-affine displacement ($\mathbf{u}_{na}^i$) is defined as the
difference between the mean displacement of a bead $i$ during time
$\delta t$ ($\mathbf{r}^i(t+\delta
t)-\mathbf{r}^i(t)$), and the mean displacement it would experience if
the deformation were    perfectly affine, \emph{i.e} homogeneous at all
scales,
\begin{equation}
\mathbf{u}^i_{na}(t)=\mathbf{r}^i(t+\delta t)-\mathbf{r}^i(t) -\dot{\epsilon}_{yy}(t)\delta t\ r_y^i(t) \mathbf{e}_y,
\end{equation}
where $\mathbf{r}^i(t)$ is the position of bead $i$ at time $t$,
$r_y^i(t)$ is the projection of this position along along the
$y$ axis and $\delta t$ is the time elapsed between two
configurations where the NAD is evaluated (typically 30$\tau$).

In Figure~\ref{naf_displacement.fig}(a), the cavity nucleates at
$\epsilon_{yy}=0.09$. At the same strain, the NAD exhibits a
peak. Beads that exhibit the largest NAD are those which belong to the
surface of the cavity (see Figure~\ref{naf_displacement.fig}(c)).
Figure \ref{naf_displacement.fig}(b) shows the evolution of the NAD
distribution for several deformations.  Before cavitation, increasing
the deformation shifts the distribution tail to larger NAD until the
very moment at which the growth of a cavity occurs, which is
associated with very large values of NAD (see
Figure~\ref{naf_displacement.fig}(c)). A threshold for NAD magnitude has been defined: if $|\mathbf{u}^i_{na}| > 1.5\sigma$ at the yielding point, the bead $i$ is said to belong to the cavity surface. This threshold is used to identify the ``cavity beads'' in order to follow some of their local properties. The position of the cavity is defined as the centre of mass of these ``cavity beads''. After cavitation, the distribution returns to a narrower shape. Note that the NAD distribution broadens even before the stress drop in the stress-strain curve, due to the nucleation of the cavity. In the following sections, NAD will be used as a quantitative tool for investigating the possible correlations  with other microstructural or mechanical properties, such as Voronoi volume, hydrostatic stress, local density and local moduli. 

\begin{figure}
\includegraphics[width=\textwidth]{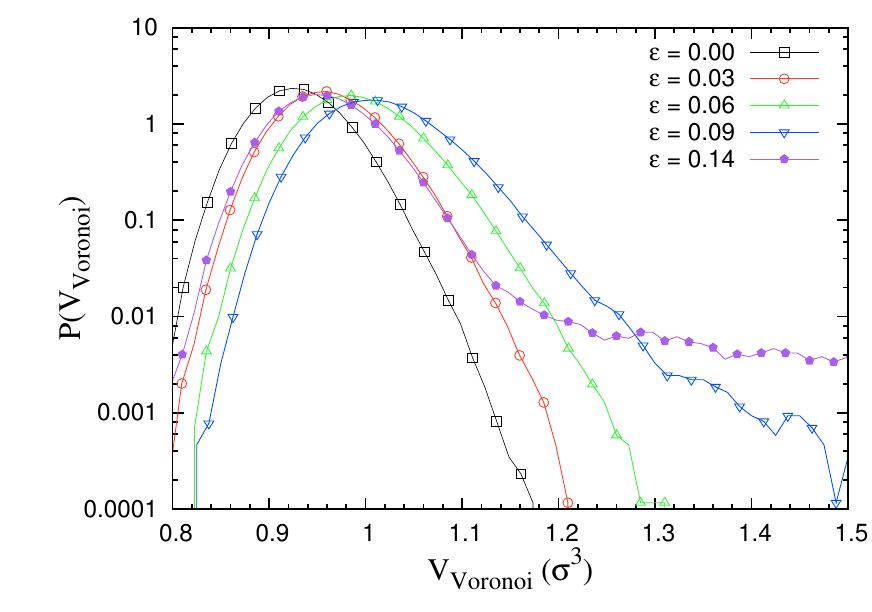}
\caption{Voronoi volume distributions of configurations extracted over
  the deformation trajectory. Cavitation can be clearly seen in the
  tail of distribution. After cavitation, the distribution reverts to
  a narrower shape.}
\label{voro_vol_distribution.fig}
\end{figure}

\section{Microstructural causes and precursors of cavitation}
\label{CausesOfCavitation.sec}
In this section, we will attempt to correlate NAD fluctuations with
some local properties measured at the scale of a single ``atom'' (
Voronoi volume and stress per atom), and properties averaged on the
scale of a few particle diameters (chain end density and bulk
modulus).
\subsection{Vorono\"{i} volume fluctuations}
\label{VoronoiVolumeFluctuations.subsec}
The concept of free volume has been extensively used to explain many
specific properties of supercooled liquids and glasses. Free volume is
defined as the volume in excess compared to an ideal disordered atomic
configuration of maximum density. One of the simplest way to compute
free volume on a local scale (and to avoid the ambiguity of the above
definition) is the Voronoi tessellation, which uniquely assigns a
polygonal volume to each bead, formed by intersecting the planes
bisecting the lines between different bead centres. In order to
determine whether local fluctuations of free volume (or Voronoi
volume) favour the nucleation of a cavity, we used the voro++ routine
to calculate the volume associated to each bead \footnote{See {\tt
    http://math.lbl.gov/voro++/} and ref.~\cite{Rycroft06}, where a
  very early version of this code was used.}.

\paragraph{Voronoi volume and deformation level.}
Figure \ref{voro_vol_distribution.fig} shows the effect of the
deformation on the Voronoi volume distribution. Increasing the
deformation will increase almost homogeneously the free volume until
cavitation takes place. During and after cavitation, the Voronoi
volume distribution exhibits a significant tail representing the beads
belonging to cavity walls.  Note that after cavitation, the
distribution relaxes to a narrower shape. Therefore, the cavitation
process can be seen as an event, which \emph{localizes} or
\emph{precipitates} the excess of free volume introduced by
deformation.

\begin{figure}
\includegraphics[width=\textwidth]{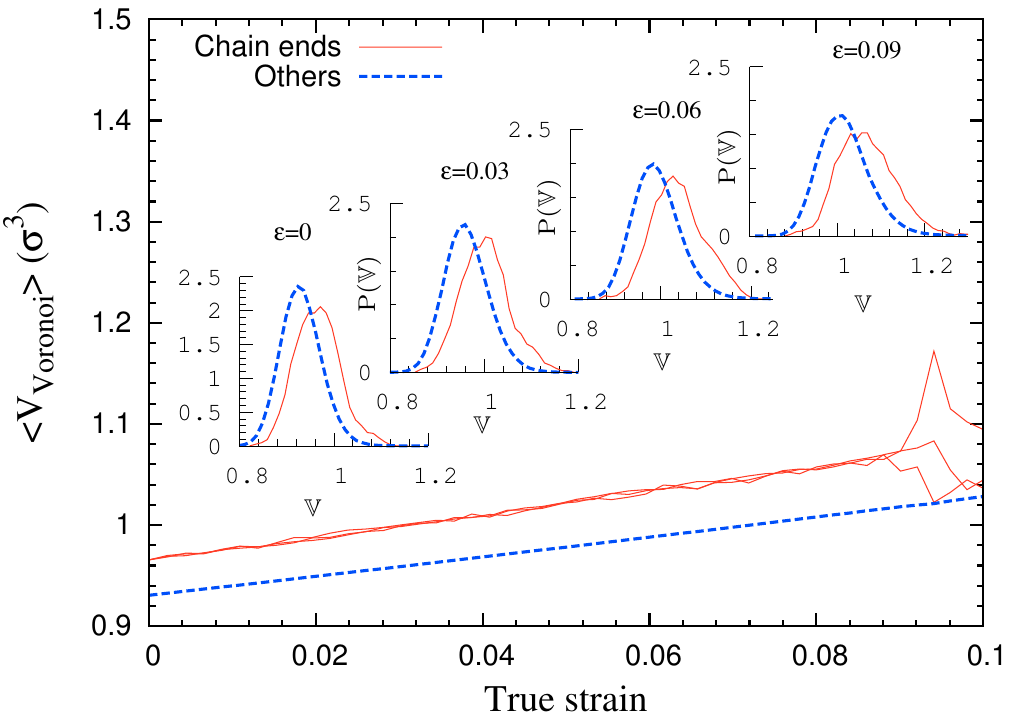}
\caption{Evolution of the mean Voronoi volume of chain ends and
  other beads.
Inset: Voronoi volume distributions during the course of deformation. No
  correlation can be observed between chain ends and Voronoi volume
  variation during tensile test.}
\label{voro_vol_distribution_focntionality.fig}
\end{figure}

\paragraph{Voronoi volume and beads functionality.}
Figure \ref{voro_vol_distribution_focntionality.fig} compares the mean
Voronoi volume evolution of both regular beads and chain ends. It
can be seen that chain ends exhibit a larger Voronoi volume, which
is not surprising since, by construction, covalent and Lennard-Jones
bonds have their energy minimum at 0.9$\sigma$ and 1.12$\sigma$,
respectively. Note that when cavitation occurs, the mean Voronoi
volume of chain ends becomes very noisy due to statistical
limitations. The insets of
Figure~\ref{voro_vol_distribution_focntionality.fig} show that the
Voronoi volume distributions have a Gaussian shape, which shows the
presence of low Voronoi volumes (much lower than the volume of an
ideal disordered configuration). This calls into question the very
concept of free volume, which is defined as that part of the atomic volume that can be redistributed throughout the system without change in energy~\cite{Turnbull61,Turnbull70}, \emph{i.e.} the volume of an ideal disordered configuration. These points of extremely low volume could be related to the \emph{constriction points}
introduced by Stachurski~\cite{Stachurski03} and, to a larger extent,
to the \emph{quasi-point defects} of Perez~\cite{Perez98}, which
represent points of high fluctuation of free energy.

\paragraph{Voronoi volume and non-affine displacement.}
Within the free volume approach, deformation induced relaxations
are supposed to be correlated with the available free volume. Zones of
larger free volume will therefore deform, changing the potential
energy landscape and providing more free volume to zones of initially
larger free volume. This explanation is often proposed to describe the
formation of mechanical instabilities such as cavitation or shear
bands. Motivated by these ideas, the search for a relationship between
the magnitude of the NAD and the Voronoi volume becomes relevant.

\begin{figure}
\includegraphics[width=\textwidth]{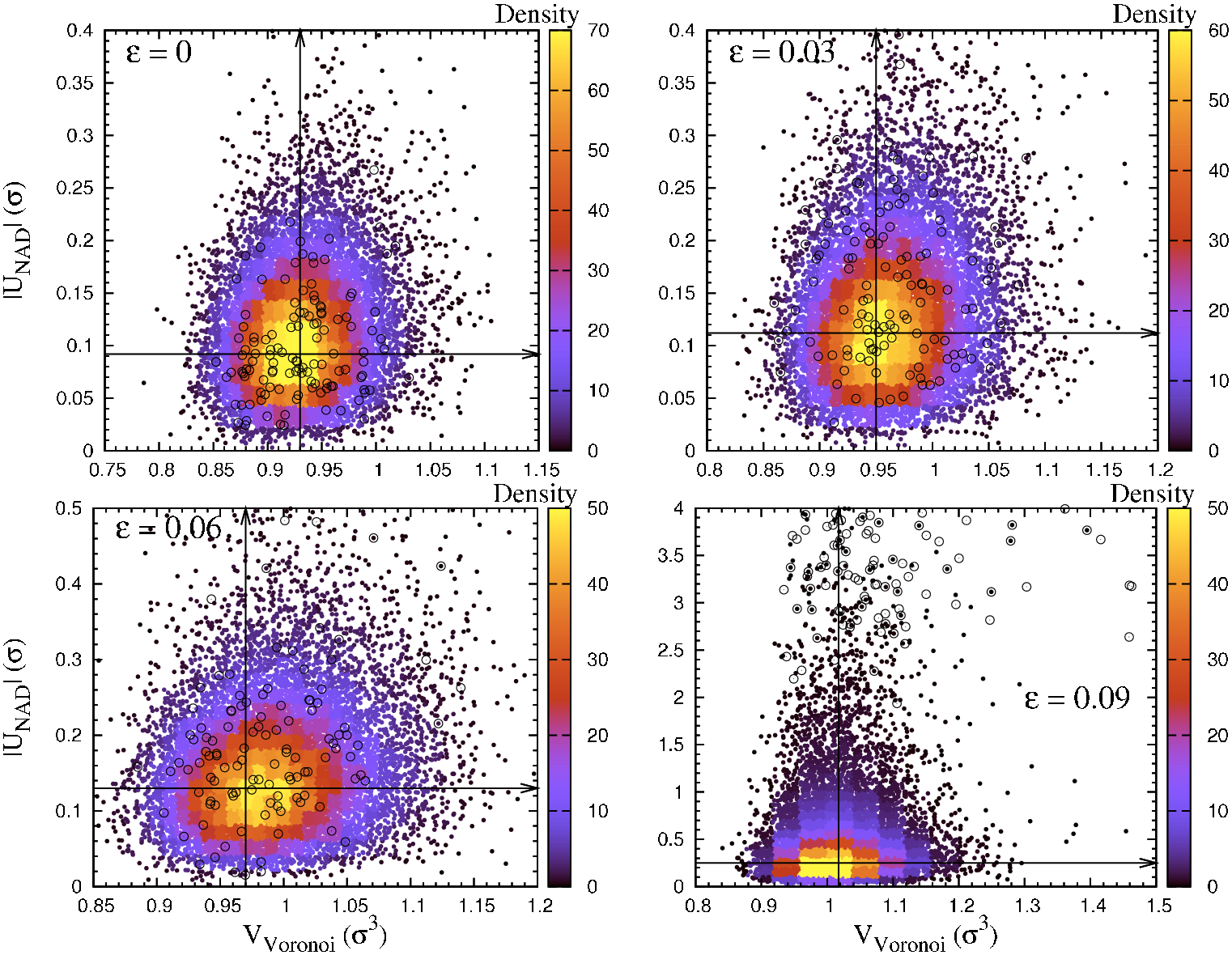}
\caption{(Color online) NAD magnitude of beads plotted against their Voronoi volume
  at several strains $\epsilon$=0, 0.03, 0.06 and 0.09
  (cavitation). Arrows describe the higher density value of each variable.
   No correlation was found between these variables for
  all strains. ``Cavity cluster beads'' are marked by the symbol
  $\circ$: no remarkable trend can be distinguished, except during
  cavitation, where they exhibit larger NAD and slightly larger
  Voronoi volume.}
\label{correlation_naf_VV_distribution.fig}
\end{figure}

Figure~\ref{correlation_naf_VV_distribution.fig} shows a scatter plot
obtained during deformation, where the magnitude of the NAD and the
Voronoi volume were taken as variables. This scatter plot does not
show a clear tendency for a correlation between NAD and Voronoi
volume. Free volume represents the potential space for motion but it
can not be seen as being a causal factor of the NAD and cavitation.
The ``cavity cluster'' beads are also shown in this plot. In both
cases, the points are distributed randomly and no noticeable trend was
found, except during cavitation, where these beads exhibit larger NAD
and slightly larger Voronoi volume. This analysis (not shown in this
paper) was performed for several other temperatures ($T=0.01$ and
$T=0.1$) and no correlation was found under these conditions either.  Note that before
cavitation, the magnitude of NAD remains much less than inter-atomic
distance ($\sigma$), in other words, the deformation is purely affine.

\subsection{Stress fluctuations}
\label{StressFluctuation.subsec}

The local stress on any given bead can be obtained by dividing the
classical expression of the virial stress by the Voronoi volume
$\mathbb{V}_m$ of atom $m$ \cite{Rottler10},
\begin{equation}\label{atomic stress}
    \sigma^m_{ij}=-\frac{1}{2\mathbb{V}_m}\left(m_m v^m_i v^m_j+\sum_{n \neq m} r^{mn}_i.f^{mn}_j \right),
\end{equation}

where $v^m_i$ is the velocity $i$th component of atom $m$; $f^{mn}_i$ and $r^{mn}_i$ are the
$i$th component of force and distance between two interacting atoms $m$ and $n$,
respectively.  The first term of this equation represent the kinetic
contribution and the second one is the Cauchy stress. The hydrostatic
stress $S_{hyd}$ was calculated by computing the trace of the stress
tensor, $S_{hyd}=-(\sigma_{11}+\sigma_{22}+\sigma_{33})/3$

\begin{figure}
\includegraphics[width=\textwidth]{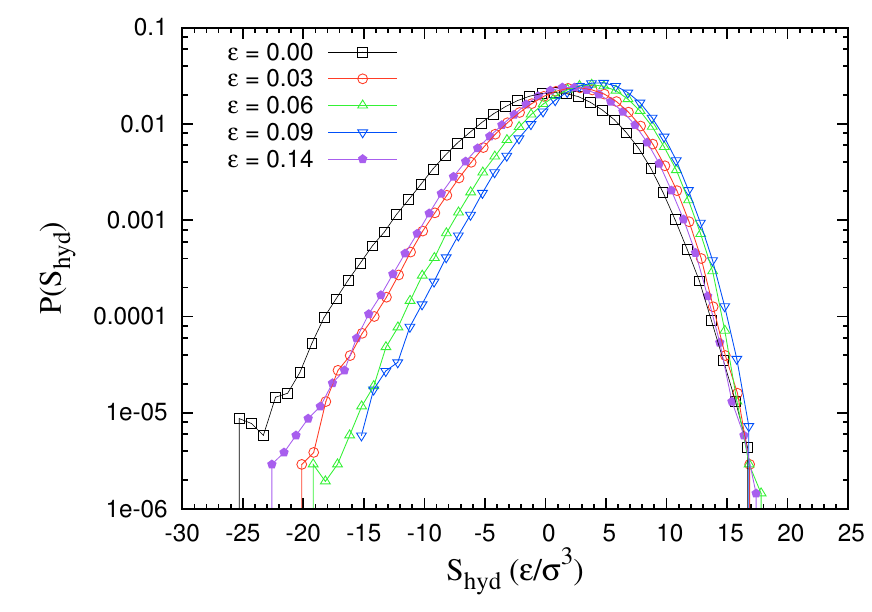}
\caption{Distribution of hydrostatic stress at different deformation levels.
The distribution narrows until the cavity opens, then broadens again.}
\label{hydro_stress_distribution.fig}
\end{figure}

Figure \ref{hydro_stress_distribution.fig} compares the distributions
of hydrostatic stresses at several strains during deformation.  In the
initial undeformed configuration, the distribution shows an
exponential tail towards negative values.  As the deformation increases, the negative values of the hydrostatic
stress are progressively relaxed, so that the distribution narrows and
becomes more symmetrical just before cavitation takes place. These
results are consistent with those of ref~\cite{Rottler10}.  After the
cavitation has occurred, large negative values of the stress are again
obtained, since the average free space is decreased as shown
previously in Figure \ref{voro_vol_distribution.fig}.

\begin{figure}
\includegraphics[width=\textwidth]{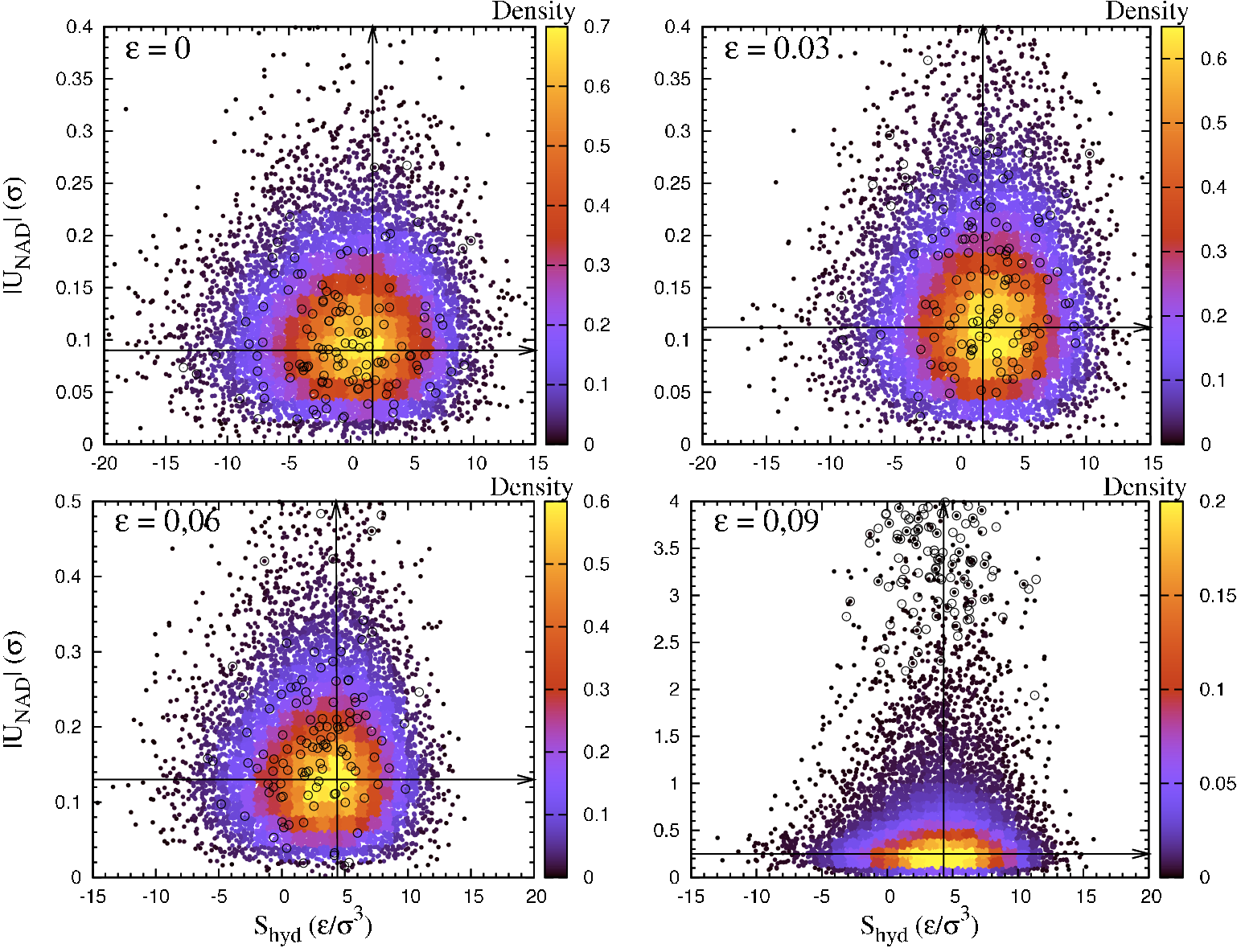}
\caption{(Color online) Scatter plot of the non affine displacement against
  hydrostatic stress shown at the same deformation as in Figure
  \ref{correlation_naf_VV_distribution.fig}. No clear trend for a
  correlation can be established. The values corresponding to the
  beads that surround the cavity labeled by the symbol $\circ$ are
  randomly dispersed, thus preventing one to identify any specific
  correlation for these beads.  }
\label{naf_hydrostatic_distribution.fig}
\end{figure}

In order to investigate the correlation between NAD and the atomic
hydrostatic stress, a scatter plot of these quantities is displayed in
Figure \ref{naf_hydrostatic_distribution.fig}.  Apparently, there is
no direct trend for a correlation between NAD and the hydrostatic
stress at the scale of individual beads.  The hydrostatic stress was
also evaluated by considering each contribution separately (pair,
bonded) at several strains, and again no correlations were found. When
specific beads (chain ends and cavity cluster) values are selected in
these scatter plots, the corresponding points appear to be a randomly
chosen subset of the total sample. This absence of correlation may
appear surprising, as the presence of a high local stress is often
expected to result in plastic deformation. 
Note however that this result is consistent with a recent study \cite{Rottler10} which 
showed no correlation between atomic stresses and shear yielding in polymers.
In an analogous way, a previous study on sheared glasses \cite{Tsamados08} also failed to find a
direct correlation between local stresses and the relevant local
plastic deformation (shear transformations in that case). 
    
It may be, however, that a more coarse grained characterization is necessary to
identify such correlations, and that the cavitation events are the
result of a local heterogeneity that extends beyond the scale of
individual beads. In order to assess this hypothesis, we describe
briefly in the next section studies performed on density fields
defined at a more coarse grained scale.

\subsection{Coarse grained densities}
The opening of a cavity under strain can be seen as a collective
event, that involves at least those atoms that will form the cavity
``skin'' at the end of the process. The corresponding mechanical
instability may therefore be the result of some density anomaly that
extends over a region larger than a single atom size or Voronoi
cell. We therefore have also explored the properties of our polymer
system on such a coarse grained scale by defining continuous fields
from the atomic positions. Various possibilities are available for
such a coarse graining procedure
\cite{Goldrisch02,Tsamados09,Detcheverry}; here we choose the simplest
one, which consists in computing the densities on a regular grid by
assigning to each grid node the atoms that belong to a fixed ``voxel''
volume around this node.  The voxel size is taken in the range
5$\sigma$ to 7$\sigma$, which was shown in similar studies
\cite{Wittmer02, Rottler10,Papakonstantopoulos08} to permit a good
description in terms of continuous fields (with about 120 monomers per
voxel) while preserving the locality and possible spatial
heterogeneity of the variables under consideration.

\begin{figure*}
\includegraphics[width=1.0\textwidth]{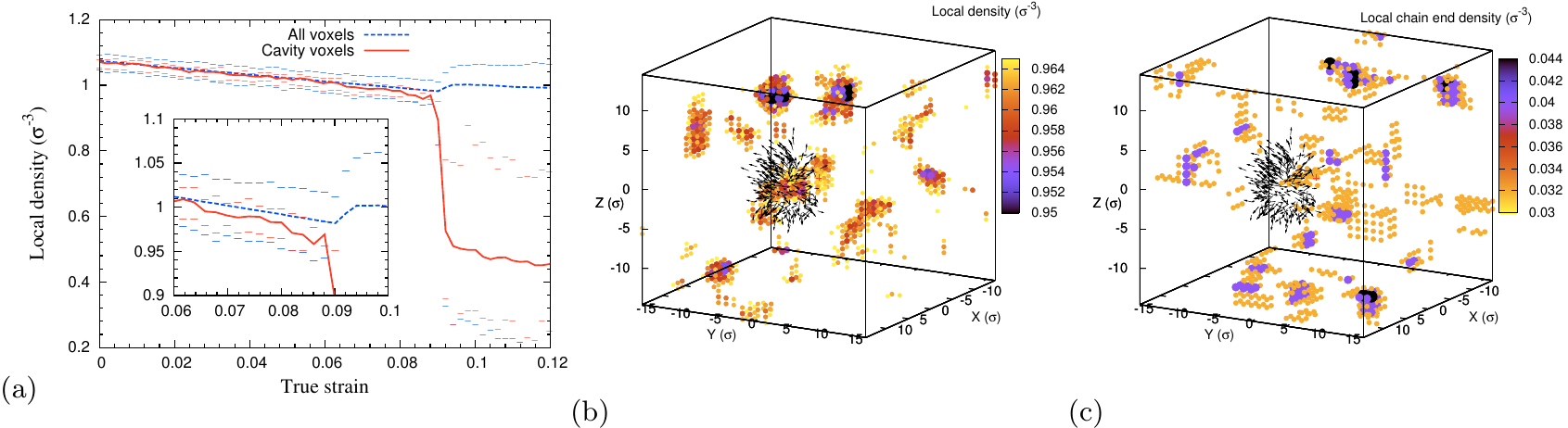}
\caption{(Color online) Evolution of mean local density calculated within cubic bins of size $5\times5\times5~\sigma^3$. The average of the overall voxels (dashed line)
  is compared with the average of voxels that are associated with the
  cavity cluster beads (solid line).  Symbols (-) denote the upper and lower values
  for each curve.  The spatial distribution of local density taken at
  $\epsilon = 0.08$ is shown in (b) and the corresponding map of chain
  end density in (c).  The arrows describe the non affine displacement
  of the ``cavity cluster beads''.  The dark spots identify the low
  density vicinities (b) and the high chain end density (c).  }
\label{Local bulk density and chain end density.fig}
\end{figure*}
We have attempted to coarse grain and to correlate with the appearance
of cavities two of the densities examined previously at the atomic
level, namely the density of chain ends and the density of monomers.
The local density field is defined as:
\begin{equation}\label{coarse grained density}
    \rho_{i}=\frac{n_i}{\sum_{j} \mathbb{V}^{j}_{i}}
\end{equation}
where $n_i$ is the number of beads within a voxel $i$, and $ \mathbb{V}^{j}_{i}$
is the Voronoi volume of the bead j that is included in the voxel $i$.
Figure \ref{Local bulk density and chain end density.fig}(a) shows the
evolution of this local density with strain, and compares the average
value with the value observed in the vicinity of the cavity.

These data show that the local density in the vicinity of the cavity
follows the mean value until a deformation of $\epsilon =0.06$.
 Although cavitation does not occur until $\epsilon = 0.09$, the
local volume begins to decrease earlier (see inset). This observed
trend can be interpreted by the fact that cavitation starts earlier
than the drop of stress in the stress-strain-curve.  This
``pre-cavitation'' behavior can be interpreted as resulting from a
dynamical equilibrium between the elastic energy and the free surface
energy of cavity with relatively small radius.  This situation remains
stable until the cavity reaches a critical radius (roughly estimated $2\sigma$), beyond which the
size of the cavity increases rapidly.  The spatial distribution of the
local density at $\epsilon = 0.08$, just before the opening the cavity
is shown in part (b) of Figure \ref{Local bulk density and chain end
  density.fig}.  The lowest density spots are far from the expected
position of the cavity, but a low density can be noticed in the cavity
vicinity.  In general, we have checked that a systematic decrease in
density prior to cavitation is specific of the points that are located
in the vicinity of the emerging cavity. Other points may display
fluctuations in their values of the density, but these fluctuations
remain uncorrelated with cavitation events. After the cavity nucleation, the low density regions that did not form cavities release their excess free volume introduced by the triaxial deformation condition. Therefore the local density of such regions return to values similar to regions that are not involved in the cavitation.
 In conclusion, local loss of density should be seen rather as a consequence than as a cause of cavity nucleation.

As was mentioned above (in section
\ref{VoronoiVolumeFluctuations.subsec}), the free volume was found to
be correlated with the bead connectivity.  Chain ends exhibit a higher
Voronoi volume compared to other monomers, and a lower density of
beads could be expected where a higher density of chain ends is
present.  We therefore define a local density of chain ends
$\rho^{C.E.}$ as
\begin{equation}\label{chain end density}
    \rho^{C.E.}_i=\frac{n^{C.E.}_i}{V_i},
\end{equation}
where $n^{C.E.}_i$ is the number of chain ends within a voxel i, and
$V_i$ is the volume of the voxel.  Figure \ref{Local bulk density and
  chain end density.fig}(c) shows that, at this level of coarse
graining, the spatial distribution of chain ends is uncorrelated with
the local density of beads and also with the cavity position. This
indicates that the modification of the packing density by the presence chain ends is insignificant.
Summarizing, the coarse grained density of beads exhibits a limited
success as a predictor for cavity formation, as its evolution can be
correlated with the formation of a cavity only shortly before the
event actually takes place. The coarse grained density of chain ends,
on the other hand, does not correlate well with the total density or
with cavitation.

\subsection{Local mechanical properties}
Our last attempt to identify a microstructural predictor for
cavitation events is inspired by previous work on simple glassy
systems under shear deformation, in which a low value of the shear
moduli was identified as a good indicator for the occurrence of the
relevant local plastic events, shear transformation zones
\cite{Tsamados09,Papakonstantopoulos08}. Here the relevant events involve a local dilatation
of the material which eventually gives rise to a cavity, and points to
the local bulk modulus as a possible predictor.

Local heterogeneity in the elastic properties of glasses is now a well
documented feature, with a number of studies having shown that the
moduli defined at intermediate scales (of the order of 10 atomic
sizes) are those of an isotropic but heterogeneous material. At such
scales, a typical glassy sample can be described as a consisting of
coexisting ``hard'' and ``soft'' regions.  This behavior is
independent of the precise method which is used to define the coarse
grained elastic constant, which may involve either the use of
statistical mechanical formulae at a local scale
\cite{Yoshimoto04,Papakonstantopoulos08}, or exploiting the linear relationship
between coarse grained stress and strain field \cite{Tsamados09}. Here
we present results for the local bulk modulus obtained from a third
approach, originally introduced by P. Sollich \emph{et.~al.}
\cite{Sollich09}, which has the advantage of being easily implemented at a reduced computational cost.
 The method can be summarized as follows: one first
defines a coarse graining volume as a fictive shape that encapsulates
a number of beads.  The shape was chosen spherical in order to reduce
any potential boundary effects, and the radius equal to 3.5 particle
diameters, consistent with the typical coarse graining scales used in
other methods \cite{Tsamados09,Papakonstantopoulos08}. The entire
sample is then deformed affinely (in this case using a uniform
dilation of all bead coordinates). After this homogeneous deformation,
all beads are kept frozen, except those contained in the coarse
graining volume which are allowed to fluctuate in a constant volume,
constant temperature molecular dynamics trajectory (here we perform a
trajectory at a rather low temperature, $T = 0.01$). The increase in
the hydrostatic stress $S^m_{hyd}$ within the spherical volume $m$ is obtained from
the virial stress formula, and the local modulus $K^m$ can be
defined by dividing this stress by the imposed increase in volumetric
strain $\vartheta^m$:

\begin{equation}\label{bulk_modulus1}
    K^{m}=\frac{d(S^m_{hyd})}{\vartheta^m}.
\end{equation}

\begin{figure}
\includegraphics[width=\textwidth]{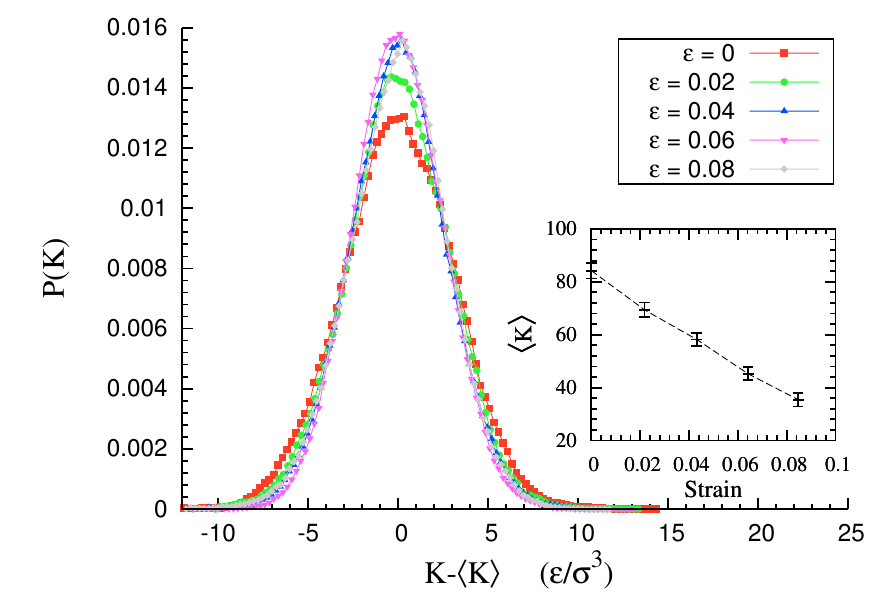}
\caption{Density distribution of local bulk modulus for the same
  specimen at several deformation levels. Distributions are shifted by
  their mean values. The inset shows the evolution of the mean
  values. Error bars are deduced from the standard deviation.  }
\label{compare_bulk_modulus distribution.fig}
\end{figure}

In order to improve the accuracy on $K^m$, it has then been averaged over a dozen expansion tests within the domain $0<\vartheta^m<10^{-5}$.  A sequence of deformation (isotropic expansion) and relaxation steps is applied over the sample, the gauge volume $V^m$ and
$S^m_{hyd}$ are measured after each step.  The expansion is
limited to a very low deformation amount since the measurement is
restricted in the elastic regime only.  Substituting the
$\vartheta^m$ by its definition $\frac{dV^m}{V^m}$ leads to another
form of equation (\ref{bulk_modulus1}):
\begin{equation}\label{bulk_modulus2}
    K^{m}\cdot\frac{dV^m}{V^m}=d(S^m_{hyd})
\end{equation}
or equivalently
\begin{equation}\label{bulk_modulus3}
    K^{m}\cdot\ln(V^m)\arrowvert_0^t=S^m_{hyd}\arrowvert_0^t,
\end{equation}
where $V^m$ and $S^m_{hyd}$ are integrated along the deformation
trajectory from $0$ to $t$. This method allows us to obtain an
accurate determination of $K^m$ in the linear regime by fitting the
data obtained for $S^m_{hyd}$=f($ln(V^m)$). This procedure was
applied along each tensile deformation trajectory, for positions of
the center of the coarse graining volume distributed on a regular
grid.

\begin{figure*}
\centering 

\includegraphics[width=0.8\textwidth]{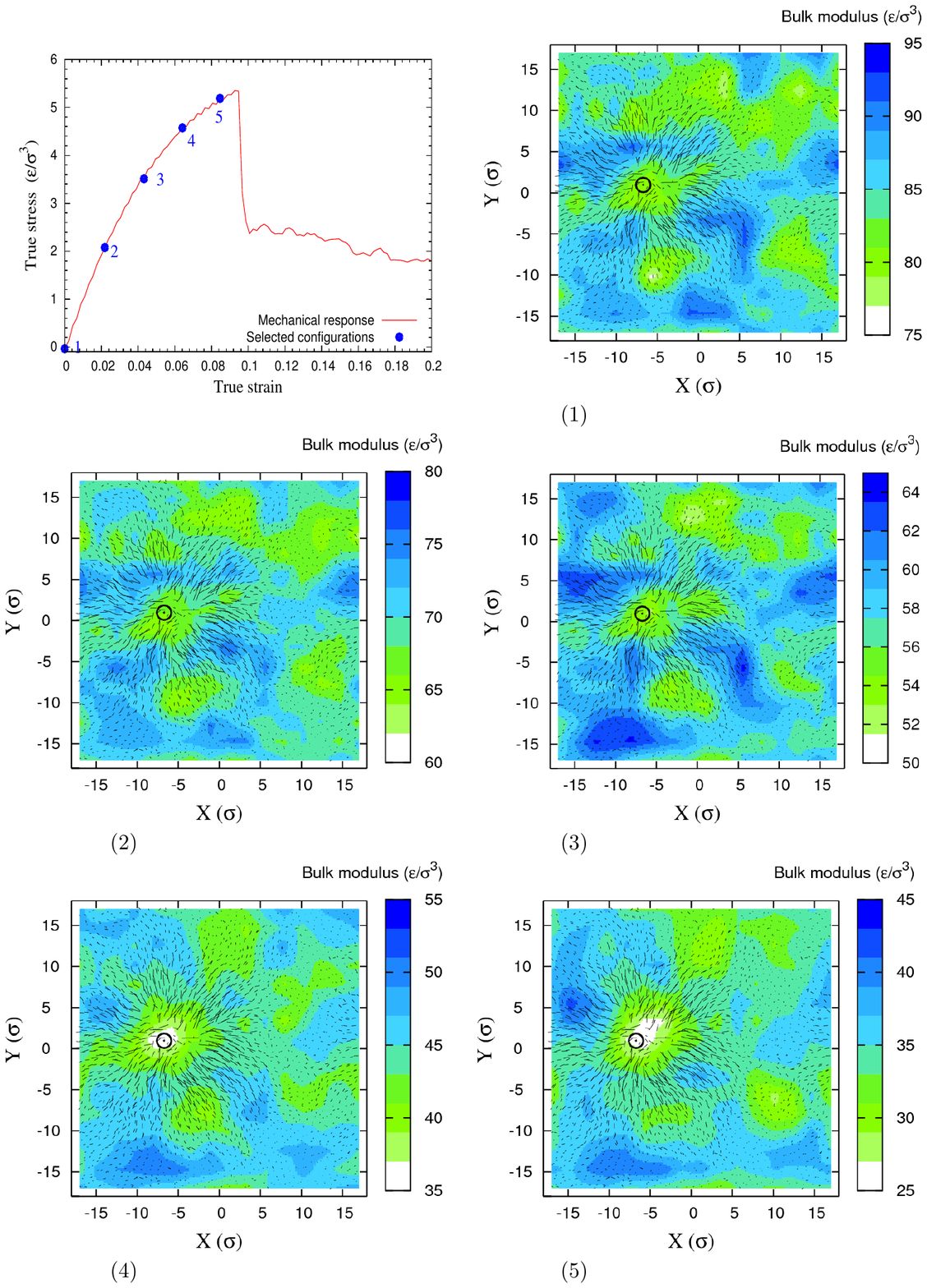}

\caption{(Color online) XY Layers taken from the spatial distribution of the bulk modulus (tensile direction is Z in this case).
The Z coordinate of all layers corresponds to the cavity position. Every layer is indexed (bottom left corner)
 referring to its position in the stress-strain curve on the upper left. Theses maps are superimposed with the nonaffine displacement at$\epsilon=0.09$
  that shows the formation of the cavity. }
\label{bulk_modulus_slices.fig}
\end{figure*}

Figure \ref{compare_bulk_modulus distribution.fig} compares the
statistical distributions of the local bulk modulus at different
strain levels.  The plotted distributions are shifted by their mean
values to facilitate comparison of their shapes.  Curves remain
symmetrical and Gaussian, whatever the applied strain before
cavitation.  As the deformation increases, the distribution will
become slightly narrower.  This behaviour is consistent with the trend
described in the previous sections, that the polymeric system tends to
homogenize its local stress under an applied deformation.  The
mean value of the local bulk modulus (see inset) decreases
continuously as the deformation increases and more free volume is
introduced in the system.

In order to investigate the spatial distribution of the local bulk
modulus, two-dimensional slices in a plane perpendicular to the
tensile direction were taken at the level at which the cavity is
observed.  Figure \ref{bulk_modulus_slices.fig} shows a sequence of
such bulk modulus cartographies that are captured along the
deformation trajectory.  Each slice corresponds to one of the blue
markers on the stress strain curve (first plot in figure
\ref{bulk_modulus_slices.fig}).  The nonaffine vectors describing the
formation of a cavity are also plotted on each map.  As can be seen,
the local bulk modulus fluctuates between high and low values at each
strain, and the position at which the cavity appears corresponds to
one of the low bulk modulus sites identified in the starting
configuration.  When the deformation increases, an extremely low value
of bulk modulus appears in the expected position of cavity, as in
slices (4) and (5).  The lower value indicated here is not only a
local minimum in the plane of the figure, but instead corresponds to
the lowest value for the entire sample.

In the light of this strong correlation between NAD and elastic
modulus, the cavitation process in glassy polymers can be described in
the following manner: The polymeric system exhibits some fluctuations
in the local elastic bulk modulus.  As the deformation progresses, the
statistical distribution of the bulk modulus changes: The mean value
decreases, but the contrast of spatial distribution is conserved.  At
relatively high strain, one of the zones that initially displayed a
low bulk modulus will reach an anomalous value, resulting in a
favorable location for the subsequent growth of a cavity.

We will now investigate whether this behavior should be described as
deterministic (the cavity systematically forms in a particular zone)
or rather statistical (the cavity forms randomly in one of the zones
with a low modulus) process. To this end, the same system was
subjected to three tensile tests with different tensile directions x,
y and z.  The positions at which cavitation takes place were recorded
and compared after each test.  

\begin{figure*}[b]
\includegraphics[width=0.9\textwidth]{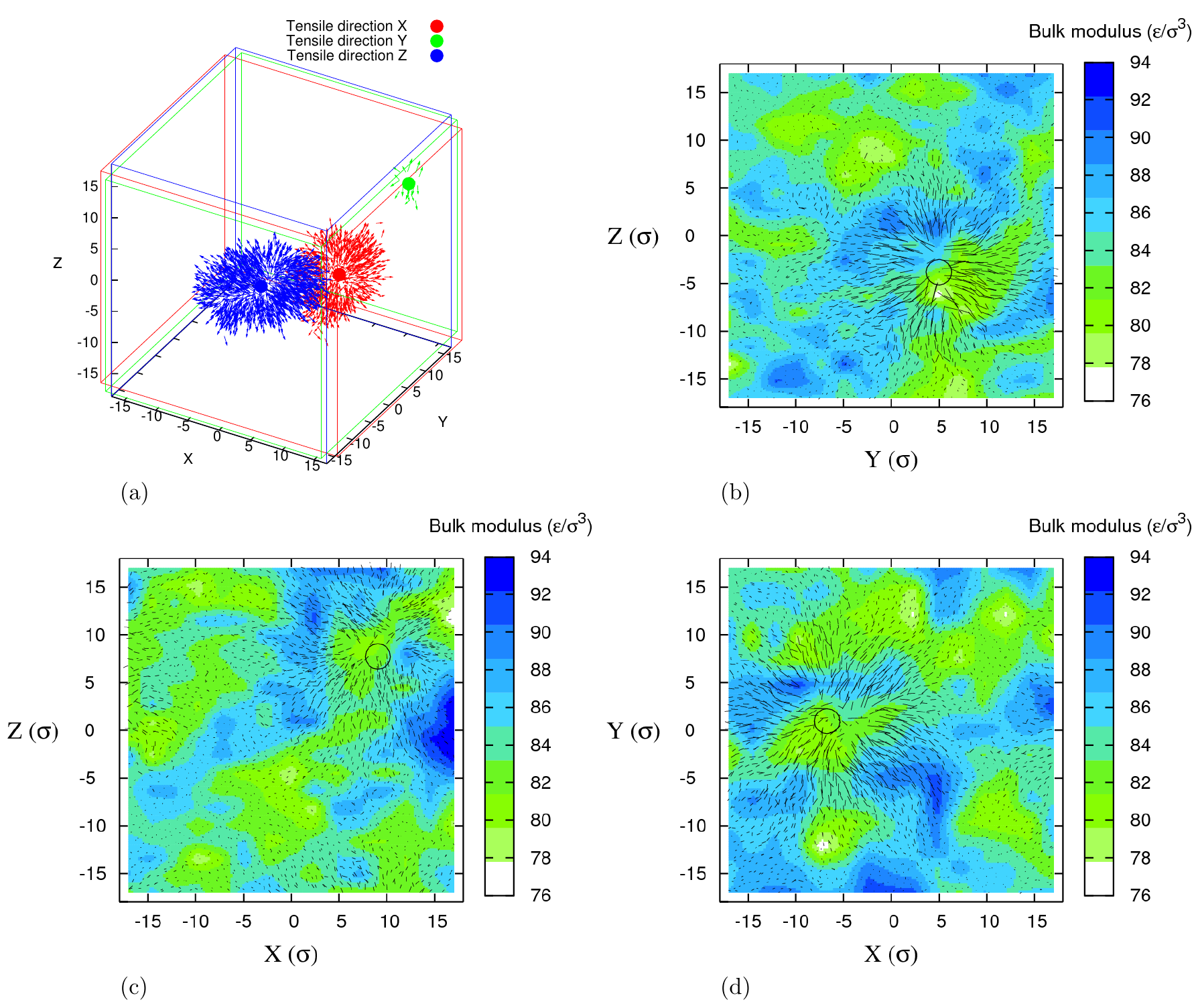}

\caption{(Color online), (a) Positions of cavities in a single sample
  after tensile tests with different straining directions. The
  vectors describing the formation of a cavity are superimposed with
  the corresponding map of local elastic bulk modulus imaged in the
  initial, undeformed, configuration. Slices (b), (c) and (d) are
  located in the corresponding position of the cavity nucleation and they are perpendicular to the
  tensile directions X, Y and Z, respectively.}
\label{cavity_position_vs_tensile_direction.fig}
\end{figure*}

Figure \ref{cavity_position_vs_tensile_direction.fig}(a) shows that, for the
same initial configuration, cavities nucleate in different zones.  The
same behaviour was also found for several systems with different
temperatures.  The cavities systematically nucleate in zones that are
characterized by a low modulus in the initial state, however the
specific site at which it is observed depends on the deformation path
and on the tensile direction.

\section{Conclusions}
\label{Conclusion.sec}
In this work the relationship between a cavitation event in a glassy
polymer undergoing a tensile test and the local properties has been
investigated with molecular dynamics simulations. Several properties
have been analyzed at two different length scales: the elementary
scale of the monomer, and a coarse graining scale of 5 to 10 particle
diameters. Independent of the scale under consideration, we find
that the density of monomers or the density of chain ends do not
correlate with the subsequent appearance of a cavity.  In contrast,
the bulk modulus in the unstrained configuration displays fluctuations
that can be directly related, in a statistical sense, to the
appearance of a cavity at large deformations. Note that very similar conclusions were reached by Toepperwein and
de Pablo in a recent study that considered both homopolymers and composite systems \cite{Toepperwein11}. This situation resembles
those observed in glassy materials under volume conserving shear,
where a weak shear modulus indicates a tendency for plastic
rearrangement.
\\
\\
\textbf{Acknowledgments:}
Computational support was provided by the Federation Lyonnaise de Calcul Haute
Performance and GENCI/CINES . The financial support from ANR project Nanomeca is
acknowledged. Part of the simulations were carried out using the
LAMMPS molecular dynamics software (http://lammps.sandia.gov).

\end{linenumbers}

\bibliographystyle{model1-num-names}

\end{document}